\documentclass{article}

\usepackage{graphicx}

\usepackage{amsmath,amsthm,amsfonts,latexsym,amscd,array}
\usepackage{bbm}
\usepackage[mathscr]{eucal}
\usepackage{amssymb}
\newtheorem{lemma}{Lemma}
\newtheorem{proposition}{Proposition}
\newtheorem{theorem}{Theorem}
\newtheorem{definition}{Definition}

\newtheorem{corollary}{Corollary}
\title{Topological Entropy}
\author{David Koslicki}
\usepackage{geometry}
\begin{document}
%\firstpage{1}

\maketitle

\begin{abstract}

Topological entropy has been one of the most difficult to implement of all the entropy-theoretic notions. This is primarily due to finite sample effects and high-dimensionality problems. In particular, topological entropy has been implemented in previous literature to conclude that entropy of exons is higher than of introns, thus implying that exons are more ``random" than introns. We define a new approximation to topological entropy free from the aforementioned difficulties. We compute its expected value and apply this definition to the intron and exon regions of the human genome to observe that as expected, the entropy of introns are significantly higher than that of exons. Though we surprisingly find that introns are less random than expected: their entropy is lower than the computed expected value.  We observe the perplexing phenomena that chromosome Y has atypically low and bi-modal entropy, possibly corresponding to random sequences (high entropy) and sequences that posses hidden structure or function (low entropy).
A Mathematica implementation is available at: http://www.math.psu.edu/koslicki/entropy.nb

\end{abstract}

\section{Introduction}

Entropy, as a measure of information content and complexity, was first introduced by Shannon (1948). Since then entropy has taken on many forms, namely topological, metric (due to Shannon), Kolmogorov-Sinai, and R\`enyi entropy. These entropies were defined for the purpose of classifying a system via some measure of complexity or simplicity. These definitions of entropy have have been applied to DNA sequences with varying levels of success. Topological entropy in particular is infrequently used due to high-dimensionality problems and finite sample effects. These issues stem from the fact that the mathematical concept of topological entropy was introduced to study \textit{infinite} length sequences. It is universally recognized that the most difficult issue in implementing entropy techniques is the convergence problems due to finite sample effects (Vinga and Almeida 2004; Kirillova 2000). A few different approaches to circumvent these problems with topological entropy and adapt it to \textit{finite} length sequences have been attempted before. For example, in Troyanskaya \textit{et al.} (2002),linguistic complexity (the fraction of total subwords to total possible subwords) is utilized to circumvent finite sample problems. This though leads to the observation that the complexity/randomness of intron regions is \textit{lower} than the complexity/randomness of exon regions. However, in Colosimo and de Luca (2000) it is found that the complexity of randomly produced sequences is \textit{higher} than that of DNA sequences, a result one would expect given the commonly held notion that intron regions of DNA are free from selective pressure and so evolve more randomly than do exon regions. Also, little has been done in the way of mathematically analyzing other finitary implementations of entropy due to most previous implementations using an entire function instead of a single value to represent entropy (thus the expected value would be very difficult to calculate)   

In this paper we focus on topological entropy, introducing a new definition that has all the desired properties of an entropy and still retains connections to information theory. This approximation, as opposed to previous implementations, is a \textit{single} number as opposed to an entire function, thus greatly speeding up the calculation time and removing high-dimensionality problems while allowing more mathematical analysis. This definition will allow the comparison of entropies of sequences of differing length, a property no other implementation of topological entropy has been able to incorporate. We will also calculate the expected value of the topological entropy to precisely draw out the connections between topological entropy and information content. We will then apply this definition to the human genome to observe that the entropy of intron regions is in fact lower than that of exon regions in the human genome as one would expect. We then provide evidence indicating that this definition of topological entropy can be used to detect sequences that are under selective pressure.

\section{Methods}

\subsection{Definitions and Preliminaries}
We restrict our attention to the alphabet $\mathscr{A}=\{A,C,T,G\}$. For a finite sequence $w$ over the alphabet $\mathscr{A}$, we use $|w|$ to denote the length of $w$. Of primary importance in the study of topological entropy is the complexity function of a sequence $w$ (finite or infinite) formed over the alphabet $\mathscr{A}$.

\begin{definition}[Complexity function]
For a given sequence $w$, the complexity function $p_w:\mathbb{N}\rightarrow \mathbb{N}$ is defined as
$$
p_w(n)=|\{u: |u|=n\ {\rm and\ u\ appears\ as\ a\ subword\ of}\ w\}|
$$
\end{definition}

That is, $p_w(n)$ represents the number of different $n$-length subwords (overlaps allowed) that appear in $w$.

Now the traditional definition of topological entropy of an \textit{infinite} word $w$ is the asymptotic exponential growth rate of the number of different subwords:

\begin{definition}
For an infinite sequence $w$ formed over the alphabet $\mathscr{A}$, the topological entropy is defined as
$$
\lim_{n\rightarrow \infty}\frac{\log_4 p_w(n)}{n}
$$
\end{definition}

\begin{figure}[hbtp!]\begin{center}
\includegraphics[width=4.5in]{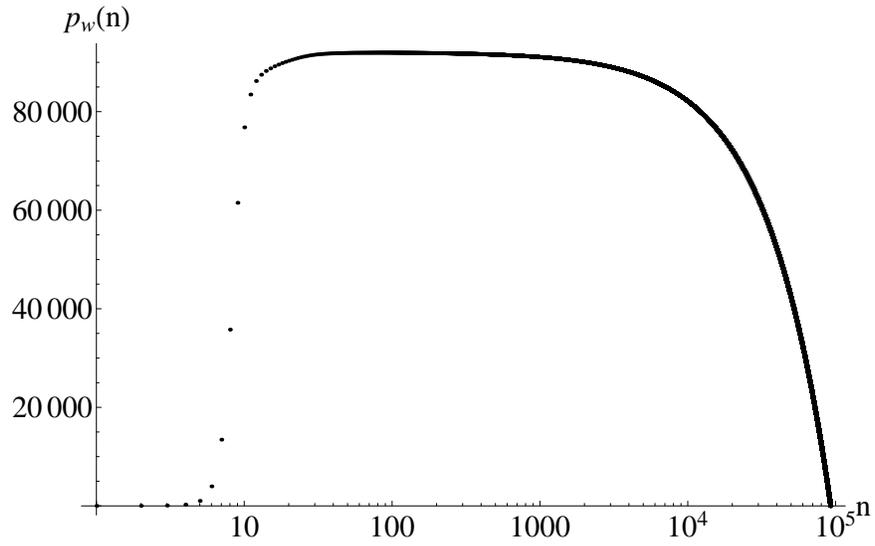}
\caption{Log-Linear Plot of the Complexity Function of the Gene ACSL4}
\label{complexity functions1}
\end{center}
\end{figure}

\begin{figure}[h!]\begin{center}
\includegraphics[width=4.5in]{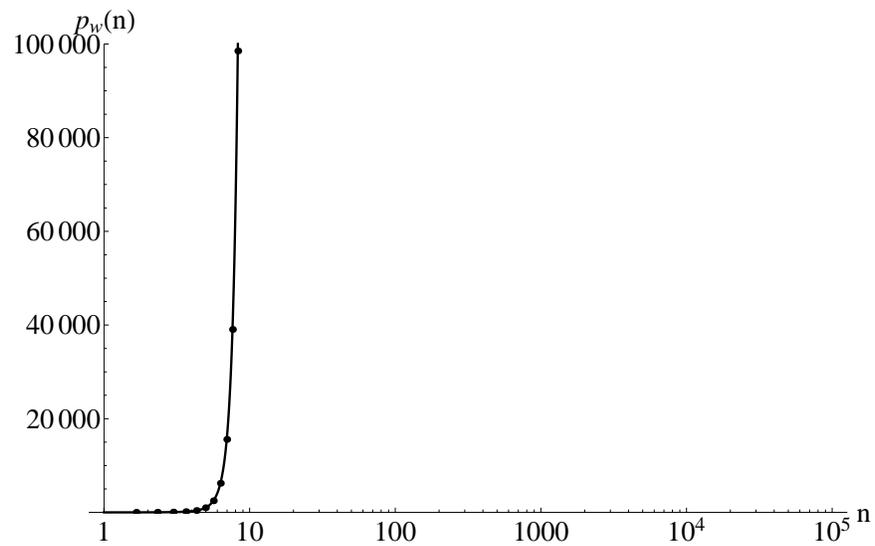}
\caption{Log-Linear Plot of the Complexity Function of a Random Infinite Sequence.}
\label{complexity functions2}
\end{center}
\end{figure}

Due to the limit in the above definition, it is easily observed that this definition will always lead to an answer of zero if applied directly to finite length sequences. This is due to the fact that the complexity function of infinite length sequences is non-decreasing, while of finite length sequences it is eventually zero. We include in figures \ref{complexity functions1} and \ref{complexity functions2} a log-linear plot of the complexity functions for the gene ACSL4 found on ChrX:108906440-108976621 (hg19) as well as for an infinite string generated by a Markov chain on four states with equal transition probabilities.

The graph of the complexity function of the gene found in figure \ref{complexity functions1} is entirely typical of the graph of a complexity function for a finite sequence as can be seen by the following proposition. The proof can be found in the nice summary by Colosimo and de Luca (2000). Note that in the following $m$ and $M$ are numbers whose calculation is straightforward.

\begin{proposition}[Shape of Complexity Function]\label{complexity function prop}
For a finite sequence $w$, there are integers $m, M$, and $N=|w|$, such that the complexity function $p_w(n)$ is strictly increasing in the interval $[0,m]$, non-decreasing in the interval $[m,M]$ and strictly decreasing in the interval $[M,N]$. In fact, for $n$ in the interval $[M,N]$ we have $p_w(n+1)-p_w(n)=-1$.
\end{proposition}

Now for a finite sequence $w$ we desire that an approximation of topological entropy $H_{top}(w)$ should have the following properties:

\begin{enumerate}
\item $0\leq H_{top}(w)\leq 1$
\item $H_{top}(w)\approx 0$ if and only if $w$ is highly repetitive (contains few subwords)
\item $H_{top}(w)\approx 1$ if and only if $w$ is highly complex (contains many subwords)
\item For different length sequences $v,w$, $H_{top}(w)$ and $H_{top}(v)$ should be comparable
\end{enumerate}

It should be noted that item 4 on this list is of utmost importance when implementing topological entropy. It is very important to normalize with respect to length since otherwise when counting the number of subwords, longer sequences will appear artificially more complex simply due to the fact that since the sequence is longer, there are more chances for subwords to show up. This explains the ``linear correlation" between sequence length and the implementations of topological entropy used in  Karamanos \textit{et al.} (2006) and Kirillova (2000). This also hints at the incomparability of the notions of entropy contained in Karamanos \textit{et al.} (2006), Colosimo and de Luca (2000), Kirillova (2000), and Schmitt and Herzel (1997).

Recall that an approximation of topological entropy should give an approximate asymptotic exponential growth rate of the number of subwords. With this and the above properties in mind, it is immediately concluded that we can disregard the values of $p_w(n)$ for $n$ in the interval $[m,N]$ mentioned in proposition \ref{complexity function prop}. In fact, as in Colosimo and de Luca (2000) the only information gained by considering $p_w(n)$ for $n$ in the interval $[m,N]$ has to do with the specific combinatorial arrangement of ``special factors" and has little to do with the complexity of a sequence.

We define the approximation to topological entropy as follows

\begin{definition}[Topological Entropy]\label{topological entropy}
Let $w$ be a finite sequence of length $|w|$, let $n$ be the unique integer such that $$4^n+n-1\leq |w|<4^{n+1}+(n+1)-1$$
Then for $w_1^{4^n+n-1}$ the first $4^n+n-1$ letters of $w$,
$$H_{top}(w):=\frac{\log_4(p_{w_1^{4^n+n-1}}(n))}{n}$$
\end{definition}

The reason for concatenating $w$ to the first $4^n+n-1$ letters is due to the following two facts whose proofs are omitted.
\begin{lemma}\label{word containment}
A sequence $w$ over the alphabet $\{A,C,T,G\}$ of length $4^n+n-1$ can contain at most $4^n$ subwords of length $n$. Conversely, if a word $w$ is to have $4^n$ subwords, it must have length at least $4^n+n-1$.
\end{lemma}
Thus if we had taken an integer $m>n$ in the above definitions and instead utilized $\frac{\log_4(p_w(m))}{m}$, $w$ would not be long enough to contain all different possible subwords.
\begin{lemma}\label{max entropy}
Say a sequence $w$ has length $4^n+n-1$ for some integer $n$, then if $w$ contains all possible subwords of length $n$ formed on the alphabet $\{A,C,T,G\}$, then $H_{top}(w)=1$
\end{lemma}
Thus if a sequence of length $4^n+n-1$ is ``as random as possible" (i.e. contains every possible subword), its topological entropy is 1, just as we would expect in the infinite sequence case. Similarly, if $w$ is ``as nonrandom as possible", that is,  if $w$ is simply the repetition of a single letter $4^n+n-1$ times, then $H_{top}(w)=0$. 

Furthermore, if we had not used concatenation in definition \ref{topological entropy}, then for a sequence $v$ such that $|v|>|w|$, the topological entropy of $v$ would on average be artificially higher due to $v$ being a longer sequence and thus has more opportunity for the appearance of subwords. Thus, by concatenating we have allowed sequences of different lengths to have comparable topological entropies.

This definition of topological entropy serves as a measure of the randomness of a sequence: the higher the entropy, the more random the sequence. The justification for this finite implementation giving an approximate characterization of randomness is given in Ornstein and Weiss (2007) in which it is shown that functions of entropy are the only finitely observable invariants of a process.

\subsection{Expected Value}
While topological entropy has been well studied for infinite sequences, very little has been done by way of mathematically analyzing topological entropy for finite sequences. This lack of analysis is most likely due to topological entropy as in the literature (Kirillova 2000;  Crochemore and Renaud 1999; Schmitt and Herzel 1997) being considered not as a single number to be associated to a DNA sequence, but rather the entire function $\frac{\log_4 p_w(n)}{n}$ is considered for \textit{every} $n$. This approach turns topological entropy (which should be just a single number associated to a DNA sequences) into a very high dimensional problem. In fact, as many dimensions as is the length of the DNA sequence under consideration. Our definition given above (definition \ref{topological entropy}) does in fact associate just a single number (instead of an entire function) to a sequence, and so is much more analytically tractable.

We now utilize the results of Gheorghiciuc and Ward (2007) to compute the expected value of the above topological entropy. This will assist us in determining what constitutes ``high" or ``low" entropy. First, we calculate the expected value of the complexity function $p_w(n)$. As is commonly assumed (Lio and Goldman 1998; Hasegawa \textit{et al.} 1985; Jukes and Cantor 1969), we now assume that DNA sequences evolve in the following way: each state in a Markov fashion independent of neighboring states. We do not assume a single model of molecular evolution, but rather just assume that there is some set of probabilities $\{\pi_A,\pi_C,\pi_T,\pi_G\}$ such that the probability of appearance of a sequence $w$ is given by the following: for $n_A$ the number of occurrences of the letter $A$ in $w$, $n_C$ the number of occurrences of the letter $C$ in $w$, etc., the probability of the sequence $w$ appearing is given by:
$$
\mathbb{P}(w)=\pi_A^{n_A}\pi_C^{n_C}\pi_T^{n_T}\pi_G^{n_G}
$$

This assumption regarding the probability of appearance of a DNA sequence is used only to procure a distribution against which we may calculate the expected number of subwords. The actual calculation of topological entropy as in definition \ref{topological entropy} does not make any such assumption about the probability of appearance.

\begin{theorem}[Expected Value of the Complexity Function]\label{expected value of the complexity function}
The expected value of the complexity function $p_w(n)$ taken over sequences of length $|w|=n+k-1$ is given by
\begin{align}\label{general expected value}
\mathbb{E}[p_w(n)]&=4^k-\sum_{w} (1-\mathbb{P}(w))^n+\mathscr{O}(n^{-\epsilon} \mu^n)
\end{align}
where the summation is over all sequences $w$ of length $n$, and $0<\epsilon<1$, $\mu<1$ (these are explicitly computed constants based on the $\pi_i$ defined above, see (Gheorghiciuc and Ward 2007)).
\end{theorem}
\begin{proof}
See (Gheorghiciuc and Ward 2007).
\end{proof}

This theorem has a particularly nice reduction when one assumes that the probability of appearance of each subletter is the same (equivalent to the the expected value being computed with a uniform distribution on the set of all sequences of a certain length).
\begin{corollary}\label{nice expected value}
Assuming that $\pi_A=\pi_C=\pi_T=\pi_G=1/4$, the expected value of complexity function taken over sequences of length $|w|=n+k-1$ is given by
\begin{align}\label{nice expected value formula}
\mathbb{E}[p_w(n)]=4^k-4^k(1-(\frac{1}{q})^k)^n+\mathscr{O}(n^{-\epsilon}\mu^k)
\end{align}
\end{corollary}

While clearly there \textit{is} a mononucleotide bias for different genomic regions and DNA sequences do not occur uniformly randomly, we do assume equal probability of appearance of each nucleotide as then the calculation of the expected number of subwords reduces in computational complexity from exponential to linear in the length of the sequence.

It is a straightforward calculation to combine formula \ref{nice expected value formula} with definition \ref{topological entropy} and compute the constants $\epsilon$ and $\mu$ as set forth in Gheorghiciuc and Ward 2007. Doing so, we obtain the following expected value for the topological entropy.

\begin{theorem}[Expected Value of Topological Entropy]\label{expected value of topological entropy}
The expected value of topological entropy taken over sequences of length $|w|=4^n+n-1$ is given by
\begin{align}\label{expected value of topological entropy formula}
\mathbb{E}[H_{top}]=\frac{\log_4(4^n-4^n(1-1/4^n)^{4^n}+\mathscr{O}((\frac{1}{\sqrt{2}}))^n)}{n}
\end{align}
\end{theorem}

We now present in table \ref{Calculated Expected Value Table} the calculated estimation of the expected value of $H_{\rm top}$ using the above formula. Keep in mind that the convergence of this calculation to the actual expected value is exponentially quick (the term $\mathscr{O}((\frac{1}{\sqrt{2}}))^n$) as $n$ increases (and so also the length of the sequence). We thus ignore the $\mathscr{O}((\frac{1}{\sqrt{2}}))^n$ term in the following calculation.
\begin{table}[!h]
\caption{Calculated Expected Value of Topological Entropy}
\label{Calculated Expected Value Table}
\begin{center}
{\small \begin{tabular}{llp{25ex}} $n$&${\small 4^n+n-1}$&Calculated Expected Value of $H_{\rm top}$\\\hline
1 & 4 & .725606\\
2 & 17 & .841242\\
3 & 66 & .890810\\
4 & 249 & .917489 \\
5 & 1028 & .933868\\
6 & 4101 & .944865\\
7 & 16390 &.952736\\
8 & 65543 & .958642\\
9 & 262152& .963237\\
10 &1048585 & .966914\\
11 & 4194315 & .969921\\
12 & 16777227 & .972428
\end{tabular}}
\end{center}
\end{table}

For comparison's sake, we present in table \ref{Sampled Expected Value Table} the sampled expected values for $n=1,\dots,9$ along with sampled standard deviations (the calculation where made by explicitly computing the topological entropy of uniformly randomly selected sequences).

\begin{table}[!h]
\caption{Sampled Expected Value and Standard Deviation of Topological Entropy}
\label{Sampled Expected Value Table}
\begin{center}
{\small \begin{tabular}{llp{10ex}p{11ex}p{15ex}} $n$&${\small 4^n+n-1}$&Sampled Expected Value of $H_{\rm top}$&Sampled Standard Deviation&Sample Size \\\hline
1 & 4 & .703583 & .184798 & 256\\
2 & 17 & .838956  & .0508640 & 300000\\
3 & 66 & .890576 & .0176785 &300000\\
4 & 249 & .917457 &.00674325 &300000\\
5 & 1028 & .933869& .0027160 &300000\\
6 & 4101 & .944861 & .00113176&300000\\
7 & 16390 &.952733 &.000486368 &300000\\
8 & 65543 &.958642 &.000212283 &300000\\
9 & 262152&.963237&.0000944814 &300000
\end{tabular}}
\end{center}
\end{table}

\begin{figure}[h!]\begin{center}
\caption{Histogram of Topological Entropy of Randomly Selected Sequences of Length $4^9+9-1=262152$}
\label{histogram of topological entropy}
\includegraphics[width=4.5in]{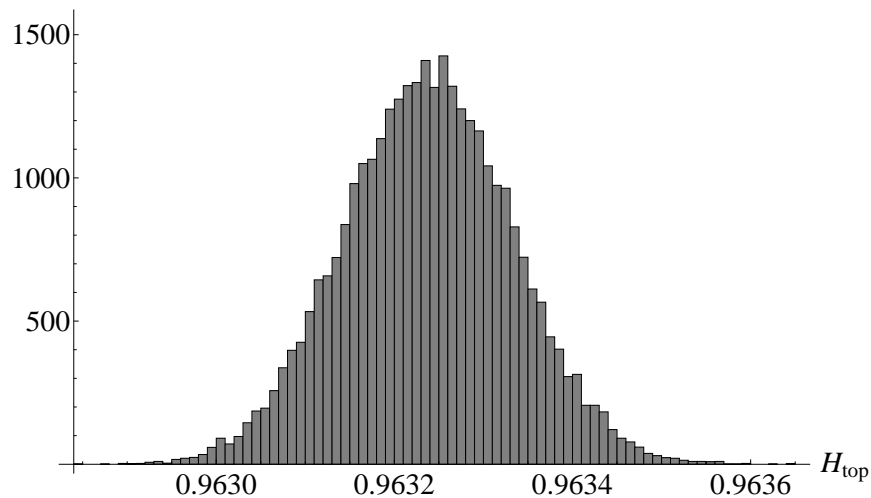}
\end{center}
\end{figure}

Summarizing this table, the topological entropy of randomly selected sequences is tightly centered around the expected value which itself is close to one. Furthermore, the distribution of topological entropy is very close to a normal distribution as can be observed from the histogram of topological entropy for sequences of length $4^9+9-1$ included in figure \ref{histogram of topological entropy}. The skewness and kurtosis are .0001996 and 2.99642 respectively.

\section{Algorithm}
An implementation of this approximation to topological entropy is available at:\\ http://www.math.psu.edu/koslicki/entropy.nb\\ We mention a few notes regarding this estimation of topological entropy. First, if a sequence $w$ in consideration has a length such that for some $n$, $4^n+n-1<|w|<4^{n+1}+n$ it will be more accurate to use a sliding window to compute the topological entropy. For example, if $|w|=16000$, we would normally concatenate this sequence to the first 4101 letters. This might misrepresent the actually topological entropy of the sequence. Accordingly, we could instead compute the average of the topological entropy of the following sequences (where $w_n^m$ means the subsequence of $w$ consisting of the $n^{\rm th}$ to $m^{\rm th}$ letters of $w$):
$$w_1^{4101},w_2^{4102},w_3^{4103},\dots, w_{11899}^{16000}$$
This is computationally intensive, so for longer sequences, one might instead choose to take non-overlapping windows, so finding the average of the topological entropy of the sequences 
$$w_1^{4101},w_{4102}^{8203},w_{8204}^{12305},\dots$$
The above website includes serial and parallel versions of the algorithm. The fastest version utilizes Nvidia CUDA GPU computing, has complexity $\mathscr{O}(n)$ for a sequence of length $n$, and takes an average of 5.2 seconds to evaluate on a DNA sequence of length 16,777,227 when using an Intel i7-950 3.6 GHz CPU and an Nvidia GTX 460 GPU.

\subsection{Comparison to Traditional Measures of Complexity}{ Other measures of DNA sequence complexity similar to this approximation of topological entropy include:  previous implementations of topological entropy (Kirillova, 2000), special factors (Colosimo and de Luca, 2000), Shannon's metric entropy (Kirillova, 2000; Farach \textit{et al.}, 1995), R\`enyi continuous entropy (Vinga and Almeida, 2004; R\`enyi, 1961), and linguistic complexity (LC) (Troyanskaya \textit{et al.}, 2002; Gabrielian and Bolshoy, 1999).
% the Lempel-Ziv approximation of Shannon's metric entropy (LZ) (Orlov and Potapov, 2004),
 
The implementation of topological entropy in Kirillova (2000) does not produce a single number representing entropy, but rather an entire sequence of values. Thus while the implementation of Kirillova (2000) does distinguish between artificial and actual DNA sequences, Kirillova notes that the implementation is hampered by high-dimensionality and finiteness problems.

In Colosimo and de Luca (2000), it is noted that the special factors approach does not differentiate between introns and exons. 

}

Note also that the convergence of our approximation of topological entropy is even faster than that of Shannon's metric entropy. Shannon's metric entropy of the sequence $u$ for the value $n$ is defined as 
$$H_{met}(u,n)=\frac{-1}{n}\sum_{w} \mu_u(w)\log(\mu_u(w))$$
where the summation is over all words of length $n$ and $\mu_u(w)$ is the probability (frequency) of the word $w$ appearing in the given sequence $u$. Thus Shannon's metric entropy requires not only the appearance of subwords, but for the actual frequency of appearance of the subwords to converge as well. As can be seen from definition \ref{topological entropy}, our notion of topological entropy does not require the use of the actual subword frequencies. So topological entropy will in general be more accurate than Shannon's metric entropy for shorter sequences. {Accordingly, the convergence issues mentioned in Farach \textit{et al.} (1995) (even with the clever Lempel-Ziv estimator) can be circumvented.}

Furthermore, it is not difficult to show (as in Blanchard \textit{et al.} (2000), Proposition 1.2.5) what is known as the \textit{Variational Principle}, that is, topological entropy dominates metric entropy: for any sequence $u$ (finite or not) and integer $n$

\begin{align}
H_{met}(u,n)\leq H_{top}(u,n) 
\end{align}
Thus topological entropy retains connections to the information theoretic interpretation of metric entropy as set forth by Shannon (1948). Since topological entropy bounds metric entropy from above:
\begin{quote}
Low topological entropy of a sequence implies that it is ``less chaotic" and is ``more structured."
\end{quote}
This connection to information theory is also an argument for the use of topological entropy over R\`enyi continuous entropy of order $\alpha$ (see Vinga and Almeida (2004) for more details).
%$$H_{\alpha}(u,n)=\frac{1}{1-\alpha} \log(\sum_w \mu_u(w))$$
%where the summation is over all words of length $n$ and $\mu_u(w)$ is the probability (frequency) of the word $w$ appearing in the given sequence $u$.
R\`enyi (1961) showed that for $\alpha \neq 1$, one cannot define conditional and mutual information functions and hence R\`enyi continuous entropy does not measure ``information content" in the usual sense. So while R\`enyi entropy does allow for the identification of statistically significant motifs (Vinga and Almeida, 2004), one cannot conclude that higher/lower R\`enyi continuous entropy for $\alpha \neq 1$ implies more/less information content or complexity in the usual sense.

Thus LC is the only other similar measurement of sequence complexity that produces a single number representing the complexity of a sequence. Like our implementation of topological entropy, the implementation of LC contained in Troyanskaya et al. (2002) also runs in linear time. A comparison of our implementation of topological entropy and LC is contained in section 4.4.

\section{Application to Exons/Introns of the Human Genome}

\subsection{Method}
We now apply our definition of topological entropy to the intron and exon regions of the human genome.

We retrieved the February 2009 GRCh37/ hg19 human genome assembly from the UCSC database and utilized Galaxy (Blankenberg \textit{et al.} 2010; Blankenberg \textit{et al.} 2007) to extract the nucleotide sequences corresponding to the introns and exons of each chromosome (including ChrX and ChrY). Now even though as argued above topological entropy converges more quickly than metric entropy, one must be careful to not use this definition of topological entropy on sequences that are too short as this would lead to significant noise. For example, the UCSC database contains exons that consist of a single base and it is meaningless to attempt to measure topological entropy of such sequences. Hence we selected the longest 100 different intron and exon sequences from each chromosome.

After ensuring that each sequence consisted only of letters from $\{A,C,T,G\}$, we then applied the approximation of topological entropy found in definition \ref{topological entropy} to the resulting sequences. For comparison's sake we also applied the approximation of topological entropy to the longest 50, 200, and 400 {sequences,} as well as {to} \textit{all} the intron and exon sequences. The salient observed features persist throughout. Though as expected, when shorter sequences are allowed, the results become noisier.

To investigate in more detail the relationship between regions under selective pressure and the value of topological entropy, we also selected each 5' and 3' UTR on chromosome Y that consisted of more than $4^3+3-1=66$ bp.

\subsection{Data}
\begin{figure}[h!]
\caption{Error bar plot of average topological entropy for the longest 100 introns and exons in each chromosome}
\label{ave bar chart 100}
\includegraphics[width=6in]{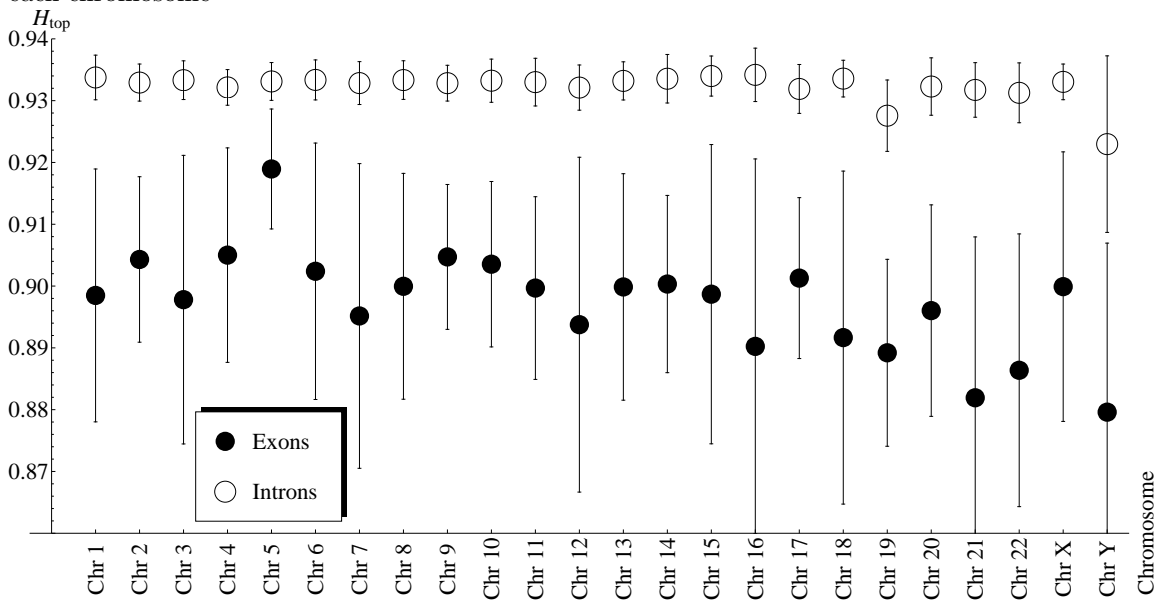}
\end{figure}

\begin{figure}[h!]
\caption{Error bar plot of chromosome Y 5' and 3' UTRs longer than 66bp long}
\label{UTRerror}
\includegraphics[width=6in]{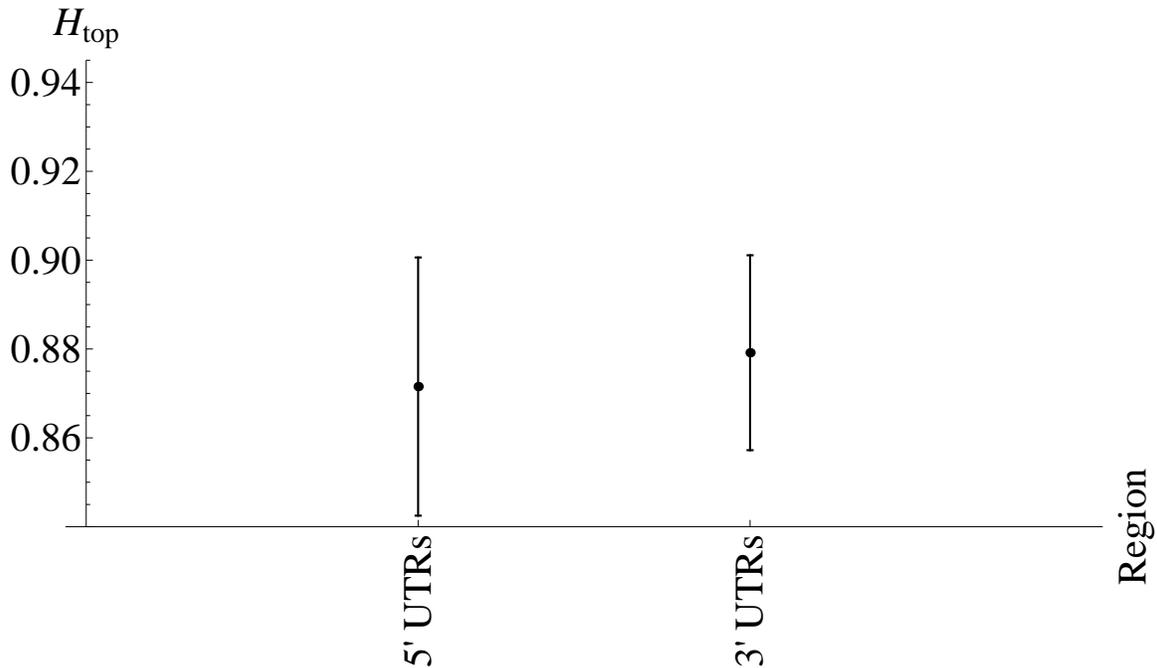}
\end{figure}

Figure \ref{ave bar chart 100} displays the error bar plot for {the} longest 100 exons and introns. The error bar plots for the longest 50, 200, and 400 {sequences,} as well as {the plot for} all the intron and exon sequences are, for brevity's sake, not shown. Figure \ref{UTRerror} displays the error bar plot for chromosome Y 5' and 3' UTRs which are longer than 66bp long.
%\newgeometry{width=12cm, left=2cm}

%\restoregeometry

\subsection{Analysis and Discussion}

We first discuss the results regarding intron and exon regions. As figure \ref{ave bar chart 100} demonstrates, the topological entropies of intron regions of the human genome are larger than the topological entropies of the exon regions. For example, the mean of the entropies of the introns on chromosome 21 is more than 11 standard deviations away from the mean of the entropy of the exons on the same chromosome. This result supports the commonly held notion that intron regions of DNA are {mostly} free from {selective} pressure and so evolve more randomly than do exon regions. We thus suggest that the observation of Karamanos \textit{et al.} (2006), Troyanskaya \textit{et al.} (2002), Mantegna \textit{et al.} (1995), and Stanley \textit{et al.} (1999) that intron entropy is \textit{smaller} than exon entropy is due to the aforementioned finite sample effects and high-dimensionality problems related to previous implementations of entropy. 

Interestingly, even though we observe that intron entropy is larger than exon entropy, the entropies of \textit{both} regions are much lower than expected (here expectation is as calculated in table \ref{Calculated Expected Value Table}). Indeed, of the longest 100 sequences, the average intron length is 180880 and the average exon length is 2059, so according to tables \ref{Calculated Expected Value Table} and \ref{Sampled Expected Value Table}, we would expect the entropies to be .966914 and .933853 respectively. We find, though, that the average entropy for introns is .9323166 and for exons is .897451. Note that the largest intron sequence entropy ($H_{top}=.943627$ for an intron of length 1.1Mbp found on chromosome 16) is significantly lower than the expected value of .969921 (at least 60 standard deviations from the expectation).{ This is not too surprising considering that the expectation as calculated in theorem \ref{expected value of topological entropy} uses the uniform distribution.} This supports the conclusion that while intron regions do evolve more randomly than exon regions, introns do not evolve uniformly randomly.

%Focusing on chromosome 5, it is particularly interesting to note that the entropy of chromosome 5 exon regions is much higher (and has a smaller standard deviation) than any other non-sex chromosome. This might partially be explained by the fact that chromosome 5 has one of the lowest gene densities in the human genome (Schmutz \textit{et al.} 2004). Having low gene density would imply that the exon regions do not exhibit a high degree of structure, and so would imply that the topological entropy would in fact be larger than a gene rich chromosome (like chromosome 19). As expected, chromosome 19 has a much lower topological entropy than chromosome 5 ($.88920\pm.01512$ versus $.91895\pm.009702$) and chromosome 19 has the highest gene density of all human chromosomes, more than double the genome wide average (Grimwood \textit{et al.} 2004).

\begin{figure}[htbp!]\begin{center}
\label{ChrYHistogram}
\caption{Histogram of topological entropy of introns in chromosome Y}
\includegraphics[width=4.5in]{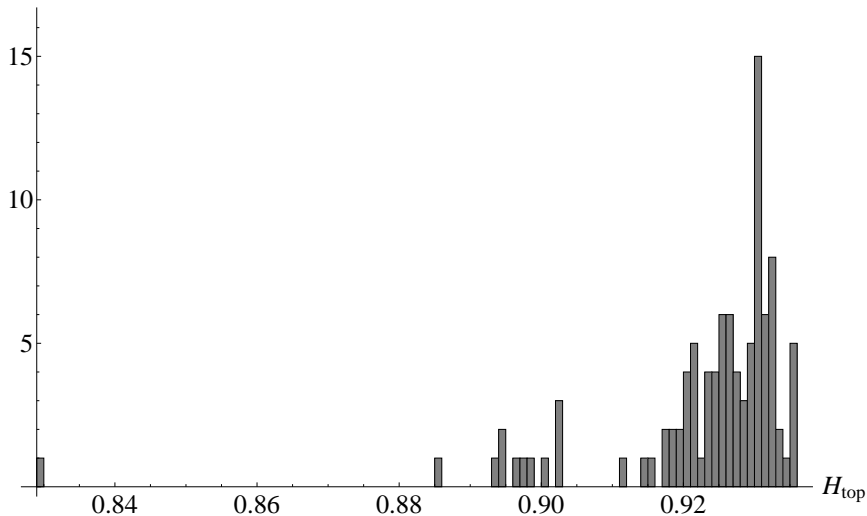}
\end{center}
\end{figure}

Note the disparity between the entropies of the sex chromosomes: The entropy of chromosome X in both intron and exon regions is significantly higher than in chromosome Y. In fact, the mean of chromosome X intron entropies is 3.5 standard deviations higher than the mean of chromosome Y intron entropies; the mean of chromosome X exon entropies is 1 standard deviation higher than the mean of chromosome Y exon entropies. Thus the X chromosome has intron and exon entropy similar to that of the autosomes, but chromosome Y has significantly differing exon and intron entropy. This is a particularly puzzling result considering {that} chromosome Y is known to have a high mutation rate and a special selection regime (Wilson and Makova 2009a; Wilson and Makova 2009b; Graves 2006), and so one would expect the entropy of chromosome Y (both intron and exon regions) to be much higher than it is. In fact, the chromosome Y introns have the lowest mean topological entropy of any intron region across the entire genome. This would suggest that the accumulation of ``junk" DNA and the massive accumulation of retrotransposable elements mentioned in Graves (2006) have some underlying function or structure. More specifically, it appears that the intron regions in chromosome Y might fall into two categories: the truly ``junk" DNA consisting of the introns with topological entropy greater than .910, and the introns that have hidden structure consisting of those sequences with entropy less than .910. We present in figure \ref{ChrYHistogram} a histogram of the topological entropy on chromosome Y demonstrating the distinction between the two categories.

Remaining on chromosome Y, we now present evidence that topological entropy can be used to detect sequences that are under selective pressure. Note that Siepel \textit{et al.} (2005) showed that both 5' and 3' UTRs are among the most conserved elements in vertebrate genomes. Thus one would expect that the topological entropy of these regions would be very low (as this is indicative of a high degree of structure). As indicated in figure \ref{UTRerror}, the entropy of both the 5' and 3' region are low in comparison to the entropy of the intron and exon regions across the autosomes. In fact the mean of the topological entropy of the 5' and 3' UTRs ($.871545\pm .0290619$ and $.879163\pm .0219371$) are lower than the mean entropy of \textit{any} intron or exon region across every chromosome. The lowest mean topological entropy for an autosome is $.927802\pm.00539$ on chromosome 19, this is more than nine standard deviations \textit{higher} than the mean of topological entropy for either the 3' or 5' UTRs. This lends support to the assertion that topological entropy can be used to detect functional regions and regions under selective constraint.
\begin{figure}[htbp!]\begin{center}
\label{UTRhistogram}
\caption{Histogram of topological entropy for 5' and 3' UTRs in chromosome Y}
\includegraphics[width=4.5in]{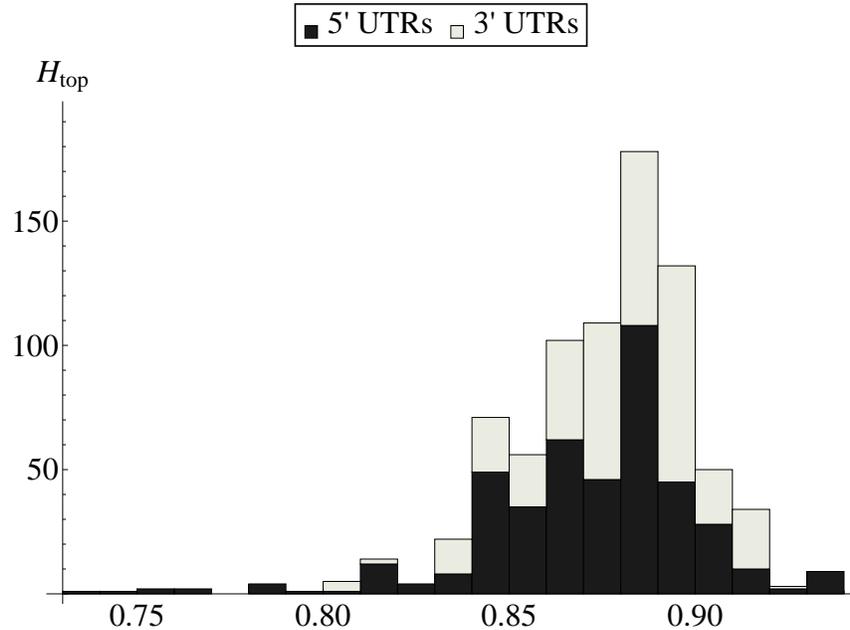}
\end{center}
\end{figure}
\subsection{Comparison to Linguistic Complexity}

\begin{figure}[h!]
\caption{Error bar plot of linguistic complexity on introns and exons using window as long as the sequence.}
\label{LC}
\includegraphics[width=6in]{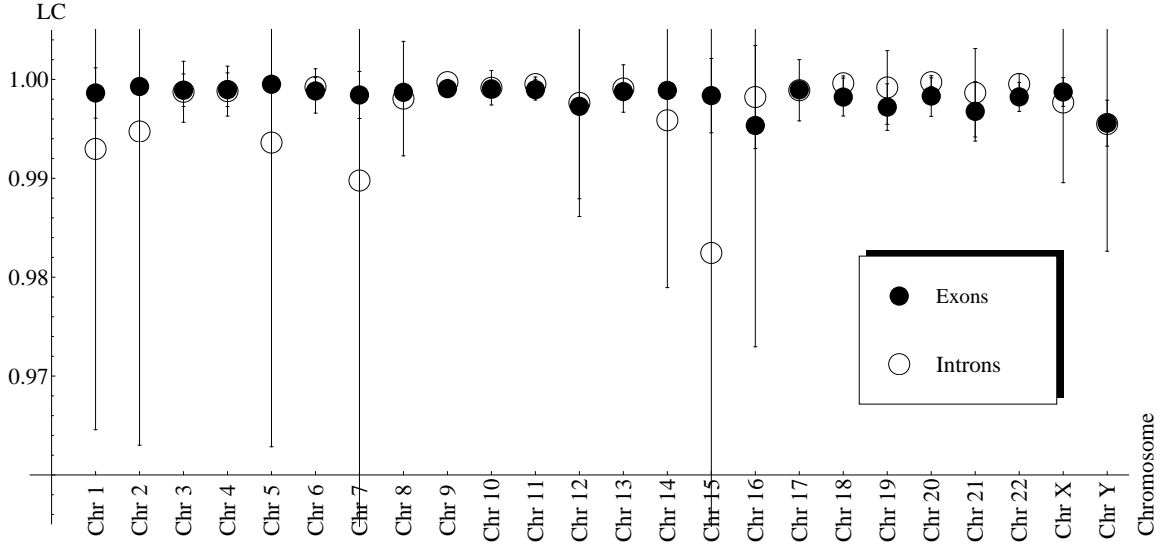}
\end{figure}

\begin{figure}[h!]
\caption{Error bar plot of linguistic complexity on introns and exons using 2000bp windows.}
\label{LC2000}
\includegraphics[width=6in]{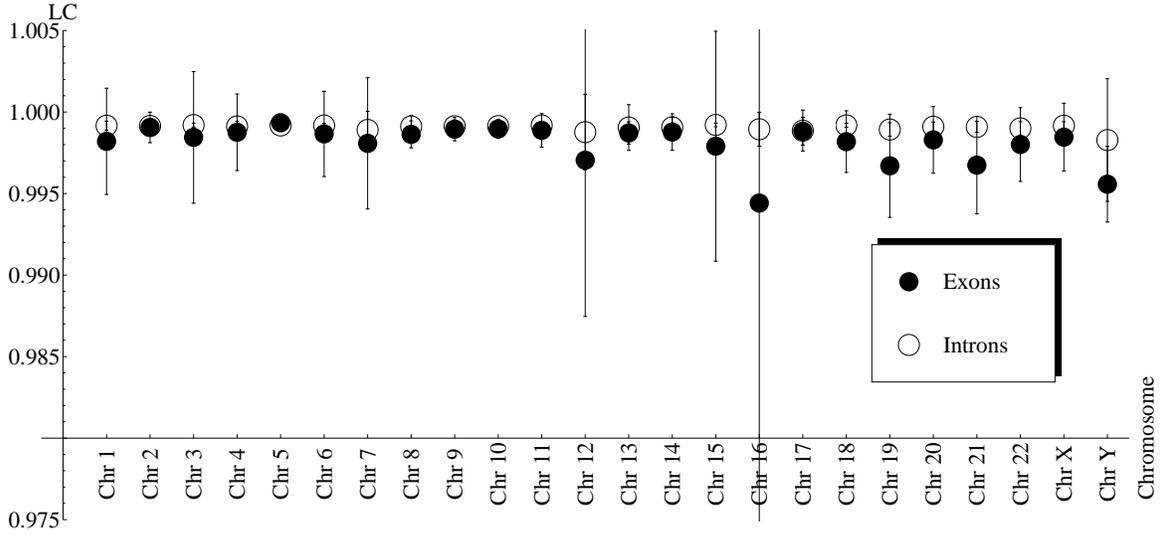}
\end{figure}

As mentioned in section 3.1, LC is the only other similar measurement of sequence complexity that produces a single number to represent the complexity of a sequence. We applied the algorithm described in Troyanskaya et al. (2002) and written by Larsson (1999) to the same data set contained in section 4.1 of this paper. To obtain directly comparable results, we used a window size as big as the given sequence is long. As can be seen in figure \ref{LC}, LC does distinguish between introns and exons to an extent, though not to the same quality of resolution as that of topological entropy (compare to figure \ref{ave bar chart 100}). For example, while topological entropy consistently measures introns as more random than exons, LC does not. This discrepancy is most likely due to linguistic complexity being effectively utilized (Troyanskaya \textit{et al.}, 2002) as a sliding window method to detect repetitive motifs, not as a holistic measure of sequence information content. So we also applied LC using a sliding window of 2000bp, taking the average value of LC on a given sequence, and then averaging on a given chromosome (see figure \ref{LC2000}). Using the sliding window, LC does give a higher value to introns than to exons (except on chromosome 5). While the separation between the LC of introns and exons becomes more pronounced, the resolution is still not nearly as clear as with topological entropy since a large amount of error persisted. The LC values amongst introns and exons are well within one standard deviation of each other across the entire genome.


\section{Conclusion}

This implementation of topological entropy is free from issues that other implementations have encountered. Namely, this definition allows for the comparison of sequences of different length and does not suffer from multi-dimensionality complications. {Since this definition supplies a single value to characterize the complexity of a sequence, it is much more capable of being mathematically analyzed.} Beyond measuring the complexity or simplicity of a sequence, we presented evidence that our approximation to topological entropy might detect functional regions and sequences free from or under {selective} constraint. The speed and simplicity of this implementation of topological entropy {makes} it very suitable for utilization in detecting regions of high/low complexity. For example, we observe the novel phenomena that {the introns} on chromosome Y {have} atypically low and bi-modal entropy, possibly corresponding to random sequences and sequences that posses hidden structure or function.

\section*{Acknowledgments}
The author would like to thank Manfred Denker, Kateryna Makova, and Francesca Chiaromonte for their assistance and fruitful discussion regarding this paper. This work was supported by the National Science Foundation [grant number DMS-1008538].

%\twocolumn

\end{document}